\documentclass{IEEEtran}
\usepackage{graphicx,amsmath,hyperref}

\begin{document}

\title{A Mathematical Model of Chaotic Attractor\\ in Tumor Growth and Decay}
\author{Tijana T. Ivancevic\thanks{Tijana.Ivancevic@unisa.edu.au}, Murk J. Bottema\thanks{murkb@csem.flinders.edu.au} and Lakhmi C. Jain\thanks{Lakhmi.Jain@unisa.edu.au}}
\date{}
\maketitle

\begin{abstract} We propose a strange-attractor model of tumor growth and metastasis.
It is a 4-dimensional spatio-temporal cancer model with strong nonlinear
couplings. Even the same type of tumor is different in every patient both in
size and appearance, as well as in temporal behavior. This is clearly a
characteristic of dynamical systems sensitive to initial conditions. The new
chaotic model of tumor growth and decay is biologically motivated.
It has been developed as a live Mathematica demonstration, see
Wolfram Demonstrator site:\\ \\ \underline
{http://demonstrations.wolfram.com/ChaoticAttractorInTumorGrowth/}
\end{abstract}


\section{Introduction}

Cancer is one of the main causes of morbidity and mortality in the
world. There are several different stages in the growth of a tumor
before it becomes so large that it causes the patient to die or
reduces permanently their quality of life. Developed countries are
investing large sums of money into cancer research in order to
find cures and improve existing treatments. In comparison to
molecular biology, cell biology, and drug delivery research,
mathematics has so far contributed relatively little to the area
\cite{Nature}.

A number of mathematical models of avascular (solid) tumor growth
were reviewed in \cite{Roose}. These were  generally divided into
continuum cell population models described by diffusion partial
differential equations (PDEs) of continuum mechanics
\cite{GaneshSprBig,GCompl} combined with chemical kinetics, and
discrete cell population models described by ordinary differential
equations (ODEs).

On the other hand, in many biological systems it is possible to
\emph{empirically} demonstrate the presence of \emph{attractors}
that operate starting from different initial conditions
\cite{TacaNODY}. Some of these attractors are points, some are
closed curves, while the others have non--integer, fractal
dimension and are termed ``strange attractors" \cite{horizons}. It
has been proposed that a prerequisite for proper simulating tumor
growth by computer is to establish whether typical tumor growth
patterns are fractal. The fractal dimension of tumor outlines was
empirically determined using the \emph{box-counting} method
\cite{frac}. In particular, fractal analysis of a breast carcinoma
was performed using a \emph{morphometric method}, which is the
box-counting method applied to the mammogram as well as to the
histologic section of a breast carcinoma \cite{frac2}.

If tumor growth is chaotic, this could explain the unreliability
of treatment and prediction of tumor evolution. More importantly,
if chaos is established, this could be used to adjust strategies
for fighting cancer using
\emph{chaos control} and/or \emph{anti-control}.\footnote{The
chaotic behavior of a system may be artificially weakened or
suppressed if it is undesirable. This concept is known as
\emph{control of chaos}. The first and the most important method
of chaos control is the so--called OGY--method, developed by
\cite{OGY}. However, in recent years, a non-traditional concept of
anti-control of chaos has emerged. Here, the non-chaotic dynamical
system is transformed into a chaotic one by small controlled
perturbation so that useful properties of a chaotic system can be
utilized \cite{StrAttr,Complexity}.}

In this paper, a plausible chaotic diffusion model of tumor
growth and decay is presented, as a combination of
theoretical modelling and empirical search.\\

\section{A reaction--diffusion cancer growth model}

Recently, a multiscale diffusion cancer-invasion model (MDCM) was
presented in
\cite{Anderson98,Anderson00,Anderson05,Anderson-Cell,Anderson06,AndersonMBE,AndersonJTB,Anderson08,Chaplain08},
which considers cellular and microenvironmental factors
simultaneously and interactively. The model was classified as
\emph{hybrid}, since a continuum deterministic model (based on a
system of reaction--diffusion chemotaxis equations) controls the
chemical and extracellular matrix (ECM) kinetics and a discrete
cellular automata-like model (based on a biased random-walk model)
controls the cell migration and interaction. The interactions of
the tumor cells, matrix--metalloproteinases (MMs),
matrix-degradative enzymes (MDEs) and oxygen are described by the
four coupled rate PDEs:
\begin{eqnarray}
\frac{\partial n}{\partial t} &=&D_{n}\nabla ^{2}n-\chi \nabla
\cdot
(n\nabla f),  \label{c1} \\
\frac{\partial f}{\partial t} &=&-\delta mf, \label{c2} \\
\frac{\partial m}{\partial t} &=&D_{m}\nabla ^{2}m+\mu n-\lambda m,
\label{c3} \\
\frac{\partial c}{\partial t} &=&D_{c}\nabla ^{2}c+\beta f-\gamma
n-\alpha c, \label{c4}
\end{eqnarray}
where $n$ denotes the tumor cell density, $f$ is the
MM--concentration, $m$ corresponds to the MDE--concentration, and
$c$ denotes the oxygen concentration. The four variables,
$n,m,f,c$, are all functions of the 3-dimensional
spatial variable $x$ and time
$t$. All equations represent diffusion except (\ref{c2}), which
shows only temporal evolution of the MM--concentration coupled to
the MDE--concentration. $D_{n}$ is the tumor cell coefficient, $D_{m}$
is the MDE coefficient and $D_{c}>0$ is the oxygen diffusion coefficient,
while $\chi ,\mu $, $\lambda $, $\delta $, $\alpha $, $\gamma $,
$\beta $ are positive constants. The other terms respectively
denote:

$\chi \nabla \cdot (n\nabla f)-$haptotaxis;

$\mu N-$production of MDE by tumor cell;

$\lambda m-$decay of MDE;

$\delta mf-$degradation of MM by MDE;

$\alpha c-$natural decay of oxygen;

$\gamma n-$oxygen uptake;  and

$\beta f-$production of oxygen by MM.

Because of its \emph{hybrid nature} (cells treated as discrete
entities and microenvironmental parameters treated as continuous
concentrations), the 4--dimensional (4D) model
(\ref{c1})--(\ref{c4}) can be directly linked to experimental
measurements of those cellular and microenvironmental parameters
recognized by cancer biologists as important in cancer invasion.
Furthermore, the fundamental unit of the model is the cell, and
the complex collective behavior of the tumor emerges as a
consequence of interactions between factors influencing the life
cycle and movement of individual cells
\cite{Anderson98,Anderson05,Anderson-Cell,AndersonJTB,Anderson08,Chaplain08}.

In order to use realistic parameter values, the system of rate
equations (\ref{c1})--(\ref{c4}) was non-dimensionalised.
The resulting 4D scaled system of rate PDEs
\cite{Anderson05} is given by
\begin{eqnarray}
\frac{\partial n}{\partial t} &=&d_{n}\nabla ^{2}n-\rho \nabla
\cdot
(n\nabla f),  \label{d1} \\
\frac{\partial f}{\partial t} &=&-\eta mf,  \label{d2} \\
\frac{\partial m}{\partial t} &=&d_{m}\nabla ^{2}m+\kappa n-\sigma
m,
\label{d3} \\
\frac{\partial c}{\partial t} &=&d_{c}\nabla ^{2}c+\nu f-\omega
n-\phi c. \label{d4}
\end{eqnarray}
In \cite{Anderson05} the values of the non-dimensional parameters
were given as:
\begin{eqnarray}
d_n = 0.0005, ~d_m = 0.0005, ~d_c = 0.5, ~\rho = 0.01, \label{par} \\~\eta
= 50,~
\kappa = 1, ~\sigma = 0, ~\nu = 0.5, ~\omega = 0.57, ~\phi =
0.025. \notag
\end{eqnarray}

The 4D hybrid PDE--model (\ref{c1})--(\ref{c4}) can be seen as a
special case of a general \emph{multi--phase tumor growth PDE}
(\cite{Roose} equation (12)),
\begin{equation}
\frac{\partial \Phi _{i}}{\partial t}+\nabla \cdot (\mathbf{v}_{i}\Phi
_{i})=\nabla \cdot (D_{i}\Phi _{i})+\lambda _{i}(\Phi _{i},C_{i})-\mu
_{i}(\Phi _{i},C_{i}),  \label{multiPh}
\end{equation}
where for phase $i$, $\Phi _{i}$ is the volume fraction ($\sum_{i}\Phi
_{i}=1 $), $\mathbf{v}_{i}$ is the velocity, $D_{i}$ is the random motility
or diffusion, $\lambda _{i}(\Phi _{i},C_{i})$ is the chemical and phase
dependent production, and $\mu _{i}(\Phi _{i},C_{i})$ is the chemical and
phase dependent degradation/death. The multi-phase model (\ref{multiPh}) has
been derived from the basic conservation equations for the different
chemical species,
\[
\frac{\partial C_{i}}{\partial t}+\nabla \cdot \mathbf{N}_{i}=P_{i},
\]
where $C_{i}$ are the concentrations of the chemical species,
subindex $a$ for oxygen, $b$ for glucose, $c$ for lactate ion, $d$
for carbon dioxide, $e$ for bicarbonate ion, $f$ for chloride ion,
and $g$ for hydrogen ion concentration; $\mathbf{N}_{i}$ is the
flux of each of the chemical species inside the tumor spheroid;
and $P_{i}$ is the net rate of consumption/production of the
chemical species both by tumor cells and due to the chemical
reactions with other species.

In this paper we will search for a temporal 3D \emph{cancer
chaotic attractor} `buried' within the 4D hybrid spatio-temporal model
(\ref{c1})--(\ref{c4}).

\section{A chaotic multi-scale cancer growth/decay model}

From the non--dimensional spatio-temporal AC model
(\ref{d1})--(\ref{d4}), discretization was performed by neglecting
all the spatial derivatives resulting in the following simple 4D
temporal dynamical system.
\begin{eqnarray}
\dot{n} &=&0,  \label{cd1} \\
\dot{f} &=&-\eta mf,  \label{cd2} \\
\dot{m} &=&\kappa n-\sigma m,  \label{cd3} \\
\dot{c} &=&\nu f-\omega n-\phi c.  \label{cd4}
\end{eqnarray}

When simulated, the temporal system (\ref{cd1})--(\ref{cd4}) with
the set of parameters (\ref{par}) exhibits a virtually linear
temporal behavior with almost no coupling between the four
concentrations that have very different quantitative values (all
phase plots between the four concentrations, not shown here, are
virtually one-dimensional). To see if a modified version of the
system (\ref{cd1})--(\ref{cd4}) could lead to a chaotic
description of tumor growth, four new parameters, $\alpha$,
$\beta$, $\gamma$, and $\delta$ were introduced. The resulting
model is
\begin{eqnarray}
\dot{n} &=&0,  \label{md1} \\
\dot{f} &=&\alpha \eta (m-f),  \label{md2} \\
\dot{m} &=&\beta \kappa n+f(\gamma -c)-m,  \label{md3} \\
\dot{c} &=&\nu fm-\omega n-\delta \phi c.  \label{md4}
\end{eqnarray}

The introduction of the parameters $(\alpha,\beta,\gamma,\delta)$
was motivated by the fact that tumor cell shape represents a visual manifestation of an underlying balance of
forces and chemical reactions \cite{Olive}. Specifically, the
parameters represent the following quantities:

 $\alpha ~~ {\rm =} $~ tumor cell volume (proliferation/non-proliferation
fraction),

 $\beta ~~ {\rm =} $~ glucose level,

 $\gamma ~~ {\rm =} $~ number of tumor cells,

 $\delta ~~ {\rm =} $~ diffusion from the surface (saturation
 level).\smallskip

A tumor is composed of proliferating ($P$) and quiescent (or
non-proliferating) ($Q$) cells. Tumor cells shift from class $P $
to class $Q $ as the tumor grows in size \cite{Kozusko}. Model
dependence on the ratio of proliferation to non-proliferation is
introduced via the first parameter, $\alpha$. The discretization
of equation (\ref{d1}) leads to cell density being modelled as a
constant in equation (\ref{cd1}). Accordingly, cell density does
not play a role in the dynamics. In (\ref{md1})--(\ref{md4}) the
cell density is re-introduced into the dynamics via the cell
number, $\gamma$. The importance of introducing $\gamma$ also
appears in connection with the cyclin-dependent kinase (Cdk)
inhibitor p27, the level and activity of which increase in
response to cell density. Levels and activity of Cdk inhibitor p27
also increase with differentiation following loss of adhesion to
the ECM \cite{Chu}.

The ability to estimate the growth pattern of an individual tumor
cell type on the basis of morphological measurements should have
general applicability in cellular investigations, cell--growth
kinetics, cell transformation and morphogenesis \cite{Castro}.

Cell spreading alone is conducive to proliferation and increases
in DNA synthesis, indicating that cell morphology is a critical
determinant of cell function, at least in the presence of optimal
growth factors and extracellular matrix (ECM) binding
\cite{Ingber}. In many cells, the changes in morphology can
stimulate cell proliferation through integrin-mediated signaling,
indicating that cell shape may govern how individual cells will
respond to chemical signals \cite{Boudreau}.
\begin{figure}[ht]
\centerline{\includegraphics[width=9.5cm]{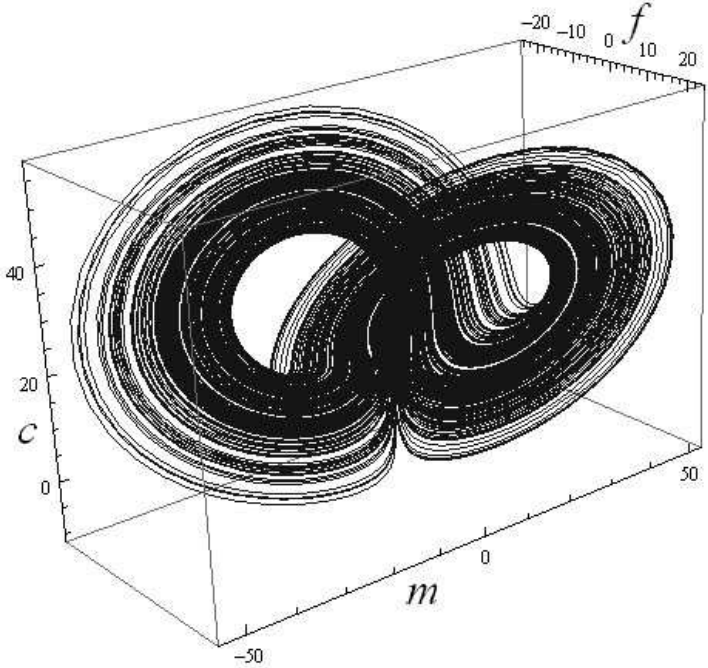}} \caption{A 3D
Lorenz-like chaotic attractor (see \cite{Lorenz,Sp}) from the
modified tumor growth model (\ref{md1})-(\ref{md4}). The attractor
effectively couples the MM--concentration $f$, the
MDE--concentration $m$, and the oxygen concentration $c$ in a
mask--like fashion.} \label{tuAtr1}
\end{figure}

Parameters $(\alpha,\beta,\gamma,\delta)$, introduced in
connection with cancer cells morphology and dynamics could also
influence the very important factor chromatin associated with
aggressive tumor phenotype and
shorter patient survival time.
For computations, the parameters were set to $\alpha $ = 0.06,
$\beta $ = 0.05, $\gamma $ = 26.5 and $\delta $ = 40. Small
variation of these chosen values would not affect the qualitative
behavior of the new temporal model (\ref{md1})--(\ref{md4}).
Simulations of (\ref{md1})--(\ref{md4}), using the same initial
conditions and the same non-dimensional parameters as before, show
chaotic behavior in the form of Lorenz-like strange attractor in
the 3D ($f-m-c$) subspace of the full 4D ($n-f-m-c$) phase-space
(see Figure \ref{tuAtr1}).

\section{A chaotic reaction--diffusion cancer growth/decay model}

The spatio-temporal system of rate PDEs corresponding to the
system in (15)--(18) provides the following multi-scale cancer
invasion model.
\begin{eqnarray}
\frac{\partial n}{\partial t} &=&d_{n}\nabla ^{2}n-\rho \nabla
\cdot
(n\nabla f),  \label{mpd1} \\
\frac{\partial f}{\partial t} &=&\alpha \eta (m-f),  \label{mpd2} \\
\frac{\partial m}{\partial t} &=&d_{m}\nabla ^{2}m+\kappa n-\sigma
m,
\label{mpd3} \\
\frac{\partial c}{\partial t} &=&d_{c}\nabla ^{2}c+\nu f-\omega
n-\phi c. \label{mpd4}
\end{eqnarray}

The new tumor--growth model (\ref{mpd1})--(\ref{mpd4}) retains all
the qualities of the original AC model (\ref{d1})--(\ref{d4}) plus
includes the temporal chaotic `butterfly'--attractor. This chaotic
behavior may be a more realistic view on the tumor growth, including
stochastic--like long--term unpredictability and
uncontrollability, as well as sensitive dependence of a tumor
growth on its initial conditions.

Based on the new tumor growth model (\ref{mpd1})--(\ref{mpd4}), we
have re-formulated the following generic concept of carcinogenesis
and metastasis (see scheme on the following page). With the model
of strongly coupled PDEs, remodelling of extracellular matrix
(ECM) causes a whole process of the movement of invading cells
with increased haptotaxis, and at the same time decreased enzyme
productions level. This can alter chromatin structure, which plays
an important role in initiating, propagating and terminating
cellular response to DNA damage \cite{Downs}. The effect of
haptotaxis in the process of cells invading could be modelled with
travelling-wave (Fisher) equation \cite{Marchant}.

The proposed model (\ref{mpd1})--(\ref{mpd4}) describes chaotic
behavior relevant to the invasion of cancer cells. As devices for controlling
metastasis/chaos we suggest the following processes:
Cellular retraining of cancer stem cells and/or activation of
positive function of cyclin-dependent kinase inhibitor p27 and/or
decreased expression of SATB1, which is correlated with aggressive
tumor phenotype in breast cancer and shorter patient survival time
\cite{Janecka,Chu,Cai,Han}.
To connect the new parameters $(\alpha,\beta,\gamma,\delta)$ to
the therapeutic regime we note the fact that significant factor of
any therapy is tumor re-growth during the rest periods between
therapy applications, which is again dependent on proliferation
fraction dynamics $\alpha $ \cite{Kozusko}.

Also, the age of the tumor may translate to different levels of
response to a standard therapy. As mentioned earlier, the number of
cells, $\gamma $, is connected with the Cdk inhibitor p27, for
which levels and activity increase in response with the cell
density $\gamma $, differentiation following loss of adhesion to
the ECM. Cdk inhibitor p27 regulates cell proliferation, cell
motility and apoptosis and is the essential element for
understanding transduction pathways in the regulation of normal
and malignant cell proliferation as well as it is new hope for
therapeutic intervention \cite{Chu}.

Introducing the parameter set $(\alpha,\beta,\gamma,\delta)$
into the model of tumor cell morphology and
function could lead to insight into the relationship between
these parameters and chromatin. By varying these parameters,
it may be possible to predict situations that result in
dangerous levels of chromatin. High levels of chromatin are
associated with cancer metastasis and some of the
most aggressive cancer types \cite{Cai,Han}.

The new temporal model (\ref{md1})--(\ref{md4}) (as well as the
corresponding spatio-temporal tumor-growth model
(\ref{mpd1})--(\ref{mpd4})) \emph{is not} sensitive to variation
of the $(\alpha,\beta,\gamma,\delta)-$values, but \emph{is}
sensitive to their corresponding places in the equations.

\section{Conclusion}

A plausible chaotic multi-scale cancer-invasion and decay model has been
presented. The new model was formulated by introducing nonlinear
coupling into the existing hybrid multiscale Anderson-Chaplain
model. The new model describes chaotic behavior, as well as
sensitive dependence of a tumor evolution on its initial
conditions. Effective cancer control is reflected in progressive reduction in
cancer mortality \cite{Janecka}. The proposed model suggests a
possible solution to carcinogenesis and metastasis, by combining
mathematical modelling with latest medical discoveries.

\end{document}